\documentclass[11pt, a4paper]{article}
\usepackage{jheparxiv}
\usepackage[latin1]{inputenc}
\usepackage{amsmath}
\usepackage{amsfonts}
\usepackage{amssymb}
\usepackage{latexsym}
\usepackage{mathrsfs}
\usepackage{graphicx}
\usepackage{color}
\usepackage{slashed}
\usepackage{twistor}
\usepackage[all]{xy}

\renewcommand{\d}{\mathrm{d}}
\renewcommand{\th}{\mathtt{h}}

\subheader{\hfill \texttt{DAMTP-2015-35}}

\title{On tree amplitudes of supersymmetric Einstein-Yang-Mills theory}

\author{Tim Adamo, Eduardo Casali, Kai A. Roehrig and David Skinner}

\affiliation{Department of Applied Mathematics \& Theoretical Physics \\
        University of Cambridge \\
        Wilberforce Road \\
        Cambridge CB3 0WA, United Kingdom}

\emailAdd{[t.adamo, e.casali, kafr2, d.b.skinner]@damtp.cam.ac.uk}

\abstract{We present a new formula for all single trace tree amplitudes in four dimensional super Yang-Mills coupled to Einstein supergravity. Like the Cachazo-He-Yuan formula, our expression is supported on solutions of the scattering equations, but with momenta written in terms of spinor helicity variables. Supersymmetry and parity are both manifest. In the pure gravity and pure Yang-Mills sectors, it reduces to the known twistor-string formulae. We show that the formula behaves correctly under factorization and sketch how these amplitudes may be obtained from a four-dimensional (ambi)twistor string.}

\notoc 

\begin{document}
 
\maketitle

\vfill

\pagebreak

\section{Introduction}

In any space-time dimension, the \emph{scattering equations} underpin the classical S-matrix of a wide variety of massless field theories by constraining external kinematic data in terms of marked points on an auxiliary Riemann sphere, $\Sigma$~\cite{Cachazo:2013gna}. Geometrically, these constraints can be compactly summarized as the requirement that a meromorphic quadratic differential $P^{2}(z)$ vanishes globally on $\Sigma$~\cite{Adamo:2013tsa}. Recently, the scattering equations have received considerable attention for their central role in the Cachazo-He-Yuan (CHY) expressions for the tree-level S-matrices of various massless bosonic field theories~\cite{Cachazo:2013hca,Cachazo:2014xea}, although they were first discovered in the context of high-energy string scattering~\cite{Fairlie:1972,Fairlie:2008dg,Gross:1987kza,Gross:1987ar}.

Even before the advent of the CHY formulae, the importance of the scattering equations was realized for field theory in four space-time dimensions~\cite{Witten:2004cp}. In four dimensions,  the spinor helicity formalism allows us to solve the on-shell condition for momentum by writing it as the product of two Weyl spinors. In addition, the simplicity of on-shell superspace makes it possible to account for arbitrary helicity external states in super-Yang-Mills theory and supergravity. Given the existence of such expressions for the pure gravity and gauge theory sectors~\cite{Roiban:2004yf,Cachazo:2012kg}, it is natural to ask if there is a generalization to the tree-level S-matrix of supergravity coupled to super-Yang-Mills theory (sEYM) in four dimensions (we consider the case where the spin one particles of the gravity multiplet remain ungauged~\cite{Bergshoeff:1985ms}). In this paper, we propose such a formula for the coupling of a single trace of gluons to gravity. It is compact, written in terms of integrals over the moduli space of a punctured Riemann sphere $\Sigma$, and supported on solutions to the scattering equations. Furthermore, it easily incorporates supersymmetry thanks to the simplicity of four-dimensional on-shell superspace, is manifestly parity invariant, produces the three-point amplitudes of the EYM action, and factorizes appropriately. 

The new formula suggests a remarkable structure in four-dimensions that would be completely obscured by a na\"ive replacement $k_i\cdot k_j\rightarrow [ij]\la ij\ra$ in the Pfaffians of the CHY formula. Remarkably, as in~\cite{Cachazo:2012kg} the $[\,,\,]$ and $\la\,,\,\ra$ factors completely decouple, appearing in separate determinants.  The previously known expressions for amplitudes in pure gauge theory or supergravity can be derived from the sphere correlation functions of certain worldsheet theories~\cite{Witten:2003nn,Berkovits:2004hg,Skinner:2013xp,Geyer:2014fka} (collectively referred to as `twistor string theories'), and our formula also appears to have a worldsheet origin.   


\section{The Formula}
\label{form}


Before describing our formula, we first point out several \textit{ab initio} constraints which any purported formula for sEYM must obey. The first of these is rather obvious: in the pure gauge theory or supergravity sectors, it must reduce to known expressions for these tree-level S-matrices. Such expressions are provided by the Roiban-Spradlin-Volovich-Witten (RSVW) formula for super-Yang-Mills theory~\cite{Roiban:2004yf,Witten:2004cp}, and the Cachazo-Skinner (CS) formula for supergravity~\cite{Cachazo:2012kg}.

A second constraint comes by considering the gravitational coupling constant $\kappa\sim\sqrt{G_{\mathrm{N}}}$. A tree-level amplitude in EYM with $n$ external gravitons and $\tau$ colour traces must be proportional to $\kappa^{n+2\tau-2}$. In particular, in the purely gravitational sector this is the usual $\kappa^{n-2}$ factor associated with (the Euler character of) a gravitational tree graph, while if there are no external gravitons and only a single colour trace, then we find $\kappa^0$ as expected for the conformally invariant Yang-Mills answer. In a connected tree, different colour traces interact by exchanging gravitons, with a gravitational coupling at each end.

In~\cite{Cachazo:2012kg} it was explained that, when written in terms of a worldsheet model, these powers of $\kappa$ must be balanced by the same number of powers of $[\, ,\,]$ or $\langle\,,\,\rangle$ brackets (interpreted as infinity twistors, representing the breaking of conformal invariance). Parity transformations exchange $[\,,\,]$ and $\la\,,\,\ra$, and so we learn that in four-dimensional EYM with $n_\pm$ external gravitons of helicity $\pm2$ and $\tau$ colour traces, we should obtain
\be\label{brcount}
\# \la\,,\,\ra = n_{-}+\tau-1\,, \qquad \# [\,,\,]=n_{+}+\tau-1 \,.
\ee 
on the worldsheet.

\medskip

A final constraint is given by a rather curious observation about the tree amplitudes of sEYM theory in four dimensions: every colour trace of external gluons must contain at least one gluon of \emph{each} helicity.\footnote{Properties of EYM closely related to this were observed in~\cite{Bern:1999bx}.} In particular, amplitudes with all negative helicity gluons and arbitrarily many positive helicity gravitons must vanish, despite the fact that this is far from obvious by standard MHV counting. To prove this, first note that each three-point interaction of the sEYM Lagrangian must contain at least one gluon of each helicity, or no gluons at all. This clear for the pure gauge theory interactions, and all three-point interactions with two gluons and one graviton arise from the term
\begin{equation*}
 \sqrt{g}\,g^{\mu\nu}\,g^{\rho\sigma}\,\partial_{[\mu}A_{\rho]}\,\partial_{[\nu}A_{\sigma]}\,,
\end{equation*}
in the Einstein-Yang-Mills action. Spinor helicity variables make it easy to see that any three-point function coming from this term with two gluons of the same helicity vanishes.

Now, consider any tree diagram in sEYM, and select all of the gluons in the diagram belonging to a particular colour trace. This identifies a unique tree sub-graph associated with the chosen trace; we may assume that all its internal edges are gluon propagators, since a graviton propagator would lead to a double trace or an external graviton contribution which can be amputated. The total number of gluons (both internal and external) of each helicity appearing in this sub-graph is equal to the number of internal edges (helicity-labelled gluon propagators) plus the number of external gluons of that helicity. The Euler characteristic of the tree sub-graph, combined with the fact that the number of gluons of each helicity at every vertex is positive, then implies the claim: the number of external gluons of each helicity in the trace must be strictly positive.

\medskip

We now turn to the expression for the tree amplitudes, beginning with the non-supersymmetric case. In this paper, we will only consider the case with $\tau\leq 1$.  External particles are specified by two spinors, $\lambda_{i\,\alpha}$ and $\tilde{\lambda}_{i\,\dot\alpha}$ for particle $i$, as well as a helicity label. We divide the external particles of each amplitude into sets of gluons and gravitons of positive or negative helicity; positive (negative) helicity gravitons form the set $\th$ ($\tth$), and positive (negative) helicity gluons form the set $\tg$ ($\ttg$). 

Let us first present the formula, and then explain its various ingredients. The tree-level amplitude with a single colour trace is given by an integral
\be\label{EYMform}
\cM_{\th,\tth}^{\tg,\ttg}\!=\!\int\!\!\frac{\mathrm{det}^{\prime}\Phi \;\mathrm{det}^{\prime}\tilde{\Phi}}{\mathrm{vol}\;\GL(2,\C)}  \, \rPT \!\! \prod_{i\in\th\cup\tg}\frac{\d t_i}{t_{i}^{|2h|-1}}\,\delta^{2}\!\left(\lambda_{i}-t_{i}\lambda(z_i)\right)
\prod_{k\in\tth\cup\ttg}\frac{\d\tilde{t}_{k}}{\tilde{t}_{k}^{|2h|-1}}\,\delta^{2}\!\left(\tilde{\lambda}_{k}-\tilde{t}_{k}\tilde{\lambda}(z_k)\right)
\ee
where $h$ denotes the helicity of the given particle (\textit{i.e.}, $h=\pm1,2$). The $\{z_i,z_k\}$ label marked points on an abstract Riemann sphere $\Sigma\cong\CP^1$ in an inhomogeneous coordinate, $z$, while the complex scaling parameters $\{t_i,\tilde{t}_k\}$ carry conformal weight, taking values in $T^{1/2}_{\Sigma\,i,k}$. The expressions $\lambda(z_i),\tilde{\lambda}(z_k)$ appearing in \eqref{EYMform} are given by
\be\label{Seqs}
\lambda(z_i)=\sum_{k\in\tth\cup\ttg}\tilde{t}_{k}\lambda_{k}\,S(i,k)\,, \qquad \tilde{\lambda}(z_k)=\sum_{j\in\th\cup\tg}t_{j}\tilde{\lambda}_j\,S(k,j)\,,
\ee
where 
\be\label{Szego}
S(z,y):=\frac{\sqrt{\d z\,\d y}}{z-y}\,,
\ee
is the Szeg\H{o} kernel at genus zero (free fermion propagator). Hence, the parameters $t_{i}$, $\tilde{t}_k$ carry opposite weight with respect to little group scalings.  

The three main ingredients in \eqref{EYMform} are the insertions $\mathrm{det}^{\prime}\Phi$, $\mathrm{det}^{\prime}\tilde{\Phi}$, and $\rPT $. $\Phi$ is a $(|\th|+ 1 )\times(|\th|+ 1)$ symmetric matrix with rows and columns corresponding to each positive helicity graviton and colour trace in the amplitude. Its entries are given by:
\be\label{Phi}
\Phi_{ij}=t_{i}t_{j}\,[i\,j]\,S(i,j) \quad  \mbox{for } i\neq j\,, \qquad \Phi_{i \tg}=\sum_{m\in\tg}t_{i}t_{m}\,[i\,m]\,S(i,m)\,,
\ee
\begin{equation*}
 \Phi_{ii}=-\sum_{j\in\th\setminus\{i\}}\Phi_{ij}- \Phi_{i \tg}\,, \qquad \Phi_{\tg\tg}= -\sum_{i\in\th}\Phi_{\tg i}\,,
\end{equation*}
where $i,j\in\th$, and $\tg$ labels the row/column for the gluon trace. It is easy to see that $\Phi$ has co-rank one, with kernel spanned by the vector $(1,1,\ldots,1)$, so its determinant vanishes. The operation $\mathrm{det}^{\prime}$ corresponds to removing any choice of row and column from a matrix and \emph{then} taking its determinant. $\mathrm{det}^{\prime}\Phi$ is easily seen to be independent of the choice of row and column removed, and is a canonically-defined non-vanishing object.

$\tilde{\Phi}$ is the parity conjugate of $\Phi$. It is a $(|\tth|+ 1)\times(|\tth|+ 1)$ matrix, with entries
\be\label{tPhi}
\tilde{\Phi}_{kl}=\tilde{t}_{k}\tilde{t}_{l}\,\la k\,l\ra\,S(k,l) \quad  \mbox{for } k\neq l\,, \qquad  \tilde{\Phi}_{k \ttg}=\sum_{m\in\ttg }\tilde{t}_{k}\tilde{t}_{m}\,\la k\,m\ra\,S(k,m)\,,
\ee
\begin{equation*}
 \tilde{\Phi}_{kk}=-\sum_{l\in\tth\setminus\{k\}}\tilde{\Phi}_{kl}- \tilde{\Phi}_{k \ttg}\,, \qquad \tilde{\Phi}_{\ttg \ttg} =  -\sum_{k\in\tth}\tilde{\Phi}_{\ttg k}\,,
\end{equation*}
where $k,l\in\tth$. This matrix also has co-rank one with kernel spanned by $(1,1,\ldots,1)$.

Finally, $\rPT $ denotes a `generalized' Parke-Taylor factor corresponding to the colour trace. If the gluon trace contains $n $ total gluons (of all helicities), then this Parke-Taylor factor is
\be\label{PT}
\rPT :=\sum_{\sigma\in S_{n }/\Z_{n }}\tr\left(\mathsf{T}^{\mathsf{a}_{\sigma(1)}}\cdots \mathsf{T}^{\mathsf{a}_{\sigma(n_\alpha)}}\right) \prod_{i=1}^{n} S(z_{\sigma(i)}, z_{\sigma(i+1)})\,,
\ee
where $\mathsf{T}^{\mathsf{a}}$ are the generators of the gauge group.

\medskip

Although $\cM_{\th,\tth}^{\tg,\ttg}$ takes the form of an integral expression, the delta functions in \eqref{EYMform} saturate all of these integrals in addition to providing overall momentum conservation. So in reality, all integrals in \eqref{EYMform} are performed algebraically against delta functions. These delta functions are a refinement of the usual scattering equations; they imply $k_{i}\cdot P(z_i)=0$ for $P_{\alpha\dot\alpha}(z)=\lambda_{\alpha}(z)\tilde{\lambda}_{\dot\alpha}(z)$~\cite{Witten:2004cp}. Note that in the special cases where $\th=\tth=\emptyset$ or $ \tg=\ttg=\emptyset $ it is equivalent to the RSVW or CS formulae respectively, presented in the guise of~\cite{Geyer:2014fka}. The reduced determinants and definitions \eqref{Phi} and \eqref{tPhi} ensure that $\cM_{\th,\tth}^{\tg,\ttg}$ is consistent with the counting of \eqref{brcount} and that the colour trace contains at least one gluon of each helicity.

\medskip

Supersymmetry can be incorporated straightforwardly, unlike in the CHY formulae, due to the simplicity of on-shell superspace in four dimensions. Extending EYM to $\cN\leq3$ supersymmetry,\footnote{We believe that the formula is also correct for $\cN=4$, provided one chooses an appropriate representation for the external gluons.} on-shell scattering states are specified by the usual two Weyl spinors $\lambda_{\alpha},\tilde{\lambda}_{\dot\alpha}$ as well as Grassmann parameters for the supermomentum, $\eta^{A}$ or $\tilde{\eta}_{A}$, for $A=1,\ldots,\cN$. Remarkably, our formula accommodates this supersymmetry with the inclusion of a single exponential factor:
\begin{multline}\label{sEYM}
 \cM_{\th,\tth}^{\tg,\ttg}=\int\frac{\mathrm{det}^{\prime}\Phi \;\mathrm{det}^{\prime}\tilde{\Phi}}{\mathrm{vol}\;\GL(2,\C)} \, ~ \rPT  ~  \exp\left(\sum_{\substack{i\in\th\cup\tg \\ k\in\tth\cup\ttg}}t_{i}\tilde{t}_{k}\,\tilde{\eta}_{A\,i}\,\eta_{k}^{A}\,S(k,i)\right) \\ 
 \times\prod_{i\in\th\cup\tg}\frac{\d t_i}{t_{i}^{|2h|-1}}\,\delta^{2}\!\left(\lambda_{i}-t_{i}\lambda(z_i)\right)\, \prod_{k\in\tth\cup\ttg}\frac{\d\tilde{t}_{k}}{\tilde{t}_{k}^{|2h|-1}}\,\delta^{2}\!\left(\tilde{\lambda}_{k}-\tilde{t}_{k}\tilde{\lambda}(z_k)\right)
\end{multline}
As usual, amplitudes for individual helicity components of the supermultiplets are read off by expanding the exponential and extracting those terms of appropriate degree in the Grassmann variables. Note that both \eqref{EYMform} and \eqref{sEYM} are manifestly parity symmetric.


\section{Justification}

In this section, we show that \eqref{sEYM} factorizes appropriately and produces the correct three-point amplitudes. 


\subsection{Three-point amplitudes}

In the helicity-based framework, all three-point amplitudes of the EYM Lagrangian are classified as MHV or $\overline{\mbox{MHV}}$, depending on whether they have one or two positive helicity legs, respectively. Since the formula \eqref{EYMform} is parity-symmetric, we only check the MHV three-point amplitudes explicitly. For Yang-Mills gauge group $\SU(N)$ there are only three such potential amplitudes:
\be\label{3ptgg}
\tth_{1}\,\tth_{2}\,\th_{3}\sim\frac{\la 1\,2\ra^{6}}{\la2\,3\ra^{2}\,\la3\,1\ra^2}\,, \qquad \ttg_{1}\,\ttg_{2}\,\tg_{3}\sim\frac{\la 1\,2\ra^{3}}{\la2\,3\ra\,\la3\,1\ra}\,, \qquad
\tg_{1}\,\ttg_{2}\,\tth_{3}\sim\frac{\la2\,3\ra^{4}}{\la1\,2\ra^2}\,,
\ee
with overall momentum conservation and any colour trace stripped off. Note that one can also write down an expression with the homogeneity of a $\ttg_{1}\ttg_{2}\th_{3}$ amplitude:
\begin{equation*}
\ttg_{1}\,\ttg_{2}\,\th_{3}\sim\frac{\la1\,2\ra^4}{\la2\,3\ra^{2}\,\la3\,1\ra^2}\,,
\end{equation*}
but this does not occur in EYM (as follows from dimensional analysis).

Since \eqref{EYMform} reduces to known expressions in the all-gluon or all-graviton sectors, it is obvious that it reproduces the first two amplitudes in \eqref{3ptgg}. So the only non-trivial calculation is to make sure that \eqref{EYMform} produces the third amplitude in \eqref{3ptgg}.  In this helicity configuration, our formula becomes
\begin{equation*}
\mathrm{tr}(\mathsf{T}^{\mathsf{a}_1}\mathsf{T}^{\mathsf{a}_2})\, \int \frac{\d \tilde{t}_{2}\,\d\tilde{t}_{3}}{\tilde{t}^{2}_{3}}\,\la2\,3\ra\,\delta^{2}\!\left(\lambda_{1}-\tilde{t}_{2}\lambda_{2}-\tilde{t}_{3}\lambda_3\right)\,\delta^{2}\!\left(\tilde{\lambda}_2+\tilde{t}_{2}\tilde{\lambda}_{1}\right)\,\delta^{2}\!\left(\tilde{\lambda}_{3}+\tilde{t}_{3}\tilde{\lambda}_{1}\right)\,,
\end{equation*}
where the positions of $z_1,z_2,z_3$ have been fixed with the $\SL(2,\C)$ freedom, the $\C^*$ freedom has been used to set $t_1=1$, and $\tilde{t}_{2,3}$ have been rescaled. The final two integrations can be done against the delta functions in a straightforward manner, leaving
\begin{equation*}
 \delta^{4}\!\left(\sum_{i=1}^3 \lambda_{i}\,\tilde{\lambda}_i\right)\,\mathrm{tr}(\mathsf{T}^{\mathsf{a}_1}\mathsf{T}^{\mathsf{a}_2})\,\frac{\la2\,3\ra^{4}}{\la1\,2\ra^2}\,,
\end{equation*}
as required.


\subsection{Factorization}

For the (non-supersymmetric) expression \eqref{EYMform} to factorize correctly, we must show that in the limit where a subset $L$ of the external momenta go on-shell
\begin{multline}\label{bfp}
\lim_{(\sum_{i\in L}\lambda_i\tilde{\lambda}_i)^2\rightarrow 0}\cM(\{\lambda_{k}\tilde{\lambda}_{k},h_k\}) = \\
\sum_{h=\pm} \int\frac{\d^{2}\lambda\,\d^{2}\tilde{\lambda}}{\mathrm{vol}\;\C^*} \cM_{L}(\{\lambda_i \tilde{\lambda}_i,h_i\}_{i\in L};\lambda\tilde{\lambda},h)\,\cM_{R}(-\lambda\tilde{\lambda},-h; \{\lambda_{j}\tilde{\lambda}_{j},h_j\}_{j\in R})\,,
\end{multline}
where the sum is over possible helicities flowing through the factorization channel, the integral is over the on-shell phase space of the intermediate state, and $R$ is the compliment of $L$. In many respects, this calculation follows similar lines to those for the RSVW and CS formulae~\cite{Skinner:2010cz,Cachazo:2012pz}.

\medskip

In terms of the Riemann surface $\Sigma$ underlying \eqref{EYMform}, the factorization limit should correspond to a degeneration of $\Sigma$ into two Riemann spheres $\Sigma_L$ and $\Sigma_R$ joined at a node. Locally, we can model this degeneration in terms of inhomogeneous coordinates by
\be\label{locmod}
(z-z_a)(z-z_b)=q\,,
\ee
where in the $q\rightarrow0$ limit the node is located at $z_a\in\Sigma_L$ and $z_{b}\in\Sigma_R$.\footnote{To be precise, the local model should read $(z_{L}-z_a)(z_{R}-z_b)=q$, where $z_L, z_R$ are appropriately chosen coordinates on $\Sigma_L,\Sigma_R$. We keep this choice of local coordinates implicit in what follows to streamline notation.} One advantage of this local model is that the behaviour of the propagator \eqref{Szego} is particularly simple near the degenerate limit:
\be\label{Szd1}
S(i,j)=\left\{\begin{array}{c}
               S(i,j) \:\:\mbox{if } z_{i},z_{j}\in\Sigma_L \:\: \mbox{or } z_{i},z_{j}\in\Sigma_R \\
               \frac{\sqrt{q}}{\sqrt{\d z_{a}\,\d z_{b}}}\,S(i,a)\,S(b,j)+O(q) \:\: \mbox{if } z_i\in\Sigma_L,\,z_j\in\Sigma_R
               \end{array}\right. \,.
\ee
This follows from the universal behaviour of Szeg\H{o} kernels on degenerate Riemann surfaces~\cite{Faybook}. 

Since the scaling parameters $t_i$, $\tilde{t}_k$ carry conformal weight, their behaviour in the degeneration limit is non-trivial. The local model \eqref{locmod} dictates that a section of $T^{1/2}_{\Sigma}$ will scale as $q^{\pm 1/4}$ in the $q\rightarrow 0$ limit, depending on which side of the degeneration the section is located.\footnote{In actuality, \eqref{locmod} only restricts a section of $T^{1/2}_{\Sigma}$ to scale as $q^{\pm\alpha/4}$ on $\Sigma_L$ and $q^{\mp(2-\alpha)/4}$ on $\Sigma_R$. We consider the symmetric case $\alpha=1$ for simplicity only.} The choice of which of $\Sigma_L$ or $\Sigma_R$ is associated with the `$+$' scaling must be summed over; below we see that this choice is associated with the helicity (positive or negative) of an intermediate particle flowing through the factorization channel.

Homogeneity of the measure on the moduli space combined with little group scaling requires that the $t_i$ and $\tilde{t}_k$ parameters scale oppositely in the degeneration limit. The behaviour of the scaling parameters is thus given by:
\begin{equation}\label{t-scaling}
t_i \sim \left\{ \begin{array}{l}  q^{\pm 1/4} \, t_i \qquad \text{for  } z_i \in \Sigma_L  \\ q^{\mp 1/4} \,  t_i \qquad \text{for  } z_i \in \Sigma_R  \end{array} \right.   \qquad \text{and} \qquad \tilde t _k \sim   \left\{ \begin{array}{l}  q^{\mp 1/4} \,  \tilde t_k \qquad \text{for  } z_k \in \Sigma_L  \\ q^{\pm 1/4}  \, \tilde t_k \qquad \text{for  } z_k \in \Sigma_R  \end{array} \right.  ~,
\end{equation}
with the choice of upper or lower sign to be summed over.

It is natural to work in a formalism where the only objects carrying conformal weight on $\Sigma$ are these parameters and the Szeg\H{o} kernels. This is accomplished by defining parameters $t_*$, $\tilde{t}_*$ (valued in $T^{1/2}_{\Sigma_L\,a}$, $T^{1/2}_{\Sigma_R\,b}$ respectively) via a single insertion of
\begin{equation}\label{const}
1 = \int \frac{\d t_{*} \d\tilde{t}_*}{\text{vol } \mathbb{C}^\ast } \, \delta \left(t_\ast \tilde t _\ast - \frac{1}{\sqrt{ \d z_{a}\,\d z_{b}  }} \right)
\end{equation}
in the amplitude \eqref{EYMform}. This allows us to re-write the behaviour of the propagator \eqref{Szego} in the attractive form:
\be\label{Szd}
S(i,j)=\left\{\begin{array}{c}
               S(i,j) \:\:\mbox{if } z_{i},z_{j}\in\Sigma_L \:\: \mbox{or } z_{i},z_{j}\in\Sigma_R \\
               \sqrt{q}\,t_{*}\tilde{t}_{*}\,S(i,a)\,S(b,j)+O(q) \:\: \mbox{if } z_i\in\Sigma_L,\,z_j\in\Sigma_R
               \end{array}\right. \,.
\ee
\emph{A priori}, the $t_*$, $\tilde{t}_*$ are just convenient dummy variables, but they will eventually become associated with an intermediate on-shell particle flowing through the degeneration.

It is also convenient to take a factor of $t_i^{-2} $ or $\tilde t_k ^{-2}$ from each graviton wave function and incorporate it into $\mathrm{det}^{\prime} \Phi$ or  $\mathrm{det}^{\prime} \tilde\Phi$, respectively. Additionally, we can choose a single gluon of each helicity, say $r\in\tg$, $s\in\ttg$, and divide the trace row and column in each of $\Phi$, $\tilde\Phi$ by the associated scaling parameter, $t_r$, $\tilde{t}_s$.

The result of these (trivial) manipulations is a transformation
\begin{equation}\label{eqn:rescaled-matrices}
\mathrm{det}^{\prime} \Phi  \, \mathrm{det}^{\prime} \tilde \Phi  ~ \prod_{i\in\th }\frac{\d t_i}{t_{i}^{3}}\, \prod_{k\in\tth }\frac{\d\tilde{t}_{k}}{\tilde{t}_{k}^{3}} ~ \to ~ t_r^2\, \tilde t_s ^2 ~  \mathrm{det}^{\prime} \Phi  \, \mathrm{det}^{\prime} \tilde \Phi ~  \prod_{i\in\th }\frac{\d t_i}{t_i}\, \prod_{k\in\tth } \frac{\d\tilde t _k}{\tilde{t}_k} \,,
\end{equation}
inside \eqref{EYMform}. Here, we abuse notation by writing $\Phi$ and $\tilde \Phi$ for the rescaled matrices with entries 
\be
\Phi_{ij}= [i\,j]\,S(i,j) \quad  \mbox{for } i\neq j\,, \qquad \Phi_{i \tg}=\sum_{m\in\tg} \frac{t_m}{t_r}\,[i\,m]\,S(i,m)\,,
\ee
\begin{equation*}
 \Phi_{ii}=-\sum_{j\in\th\setminus\{i\}} \frac{t_j}{t_i}  \, \Phi_{ij} - \frac{t_r}{t_i} \Phi_{i \tg}\,, \qquad \Phi_{\tg\tg}= -\sum_{i\in\th} \frac{t_i}{t_r} \, \Phi_{\tg i}\,,
\end{equation*}
and likewise for $\tilde \Phi$. These rescaled matrices have the (important) virtue of behaving nicely in the factorization limit. The definition of the reduced determinants is adapted appropriately for this rescaling:
\begin{align*}
\mathrm{det}^{\prime} \Phi & = \frac{|\Phi^{i}_{j}|}{t_i \, t_j } = \frac{|\Phi^{i}_{\tg}|}{t_i \, t_r } =  \frac{|\Phi^{\tg}_{\tg}|}{t_r ^2} ~, \\
 \mathrm{det}^{\prime}  \tilde \Phi & = \frac{|\tilde \Phi ^k_l|}{\tilde t_k \, \tilde  t_l } =  \frac{|\tilde \Phi ^k_{\ttg}|}{\tilde t_k \, \tilde  t_r } =\frac{|\tilde{\Phi}^{\ttg}_{\ttg}|}{\tilde  t_r ^2 } ~,
\end{align*}
where $|\Phi^{a}_b|$ denotes the determinant of $\Phi$ with row $a$ and column $b$ removed, etc. 

Note that the formula does not depend on which representative gluons we choose for rescaling the matrices. We always have at least one gluon of each helicity and the fundamental properties of the reduced determinant make it obvious that the final answer does not depend on the choice we make. Additionally, the rescaling of the matrix entries is reflected in a rescaling of the null vector; for instance, the kernel of $\Phi$ is now spanned by $(t_1,\ldots,t_{|\th|},t_r)$, which scales according to \eqref{t-scaling} in the degeneration limit.\footnote{After the various rescalings, the reduced determinants both scale like $O(\sqrt{q})$ in the $q\rightarrow 0$ limit; to make this manifest in subsequent computations, we always choose to remove a row/column corresponding to an entry of the null vector which does not tend to zero under the degeneration.} 


\medskip

Clearly, there are two different ways in which the formula \eqref{EYMform} can factorize as $q\rightarrow0$: the degeneration may or may not split the colour trace. We will show that in the former case the intermediate state corresponds to a gluon, while in the latter it will be a graviton.

Let us begin by considering the case which does not disturb the colour structure. The behaviour of \eqref{EYMform} in the $q\rightarrow0$ limit can be easily expressed by arranging the matrices $\Phi$ and $\tilde{\Phi}$ in a judicious manner. Arrange the entries of both matrices into blocks, so that the upper-left block corresponds to entries with both indices on $\Sigma_L$ and the bottom-right block corresponds to entries with both indices on $\Sigma_R$:
\begin{equation*}
 \Phi=\left(\begin{array}{c c}
             \Phi_{LL} & \Phi_{LR} \\
             \Phi_{RL} & \Phi_{RR}
            \end{array}\right)\,, 
 \qquad \tilde{\Phi}=\left(\begin{array}{c c}
                            \tilde{\Phi}_{LL} & \tilde{\Phi}_{LR} \\
                            \tilde{\Phi}_{RL} & \tilde{\Phi}_{RR}
                           \end{array}\right)\,.
\end{equation*}
Without loss of generality, we assume all gluons to remain on $\Sigma_L$. In the $q\rightarrow0$ limit, \eqref{Szd} ensures that the off-diagonal blocks of the rescaled matrices vanish at $O(\sqrt{q})$, so we must focus on what is happening in the diagonal blocks. Consider $\Phi$; clearly the entries $(\Phi_{LL})_{ij}$ and $(\Phi_{RR})_{ij}$ are unchanged on $\Sigma_L$ and $\Sigma_R$, respectively, as $q\rightarrow0$. However, the diagonal entries behave as
\begin{align*}
 (\Phi_{LL})_{ii} &=-\sum_{j\in\th_{L}\setminus\{i\}} \frac{t_j}{t_i} (\Phi_{LL})_{ij} - \frac{t_r}{t_i}(\Phi_{LL})_{i \tg} -  t_\ast \tilde t _ \ast \, \sum_{j\in\th_{R}} \frac{t_j}{t_i} [i\,j]\, S(i,a) S(b,j)+O(q)\,, \\
  (\Phi_{RR})_{ii} &=-\sum_{j\in\th_{R}\setminus\{i\}}  \frac{t_j}{t_i}  (\Phi_{RR})_{ij}  -  q \,  t_\ast \tilde t _ \ast  \!\! \sum_{j\in\th_{L}\cup\tg}  \frac{t_j}{t_i}  [i\,j]\, S(i,b) S(a,j) +O(q^{3/2})\,,
\end{align*}
where we choose the upper sign in \eqref{t-scaling} for concreteness. As the worldsheet factorizes, the particles $\th _R $ on $\Sigma_R$ form an effective particle insertion on $\Sigma _L$ which appears on the diagonal entries of $\Phi_{LL}$, while all entries of $\Phi _{RR}$ only encode the particles of $\th_R$. The interpretation is clearly that the node becomes the insertion point of a new particle with positive/negative helicity on $\Sigma_{L/R}$. (The opposite configuration follows by taking the lower sign in \eqref{t-scaling}, and the final result contains the sum over both choices.)

To make this manifest, introduce a new spinor-helicity variable $\tilde \lambda _\ast$ by inserting
\be\label{int1}
 1=\int  \d^{2}\tilde{\lambda}_{*} \,\,\delta^{2}\!\left(\tilde{\lambda}_{*}-\tilde{t}_{*}\tilde{\lambda}(z_b)\right)\,
\ee
into the amplitude \eqref{EYMform} close to the degeneration, with
\be\label{int1*}
\tilde{\lambda}(z_b):=\sum_{j\in\th_{R}}t_{j}\,\tilde{\lambda}_{j}\,S(b,j)\,.
\ee
On the support of this delta function, the diagonal entries of $\Phi$ can be rewritten as
\begin{align*}
 (\Phi_{LL})_{ii} &=-\sum_{j\in\th_{L}^\ast \setminus\{i\}} \frac{t_j}{t_i} (\Phi_{LL})_{ij} - \frac{t_r}{t_i}(\Phi_{LL})_{i \tg} + O(q)\,, \\
  (\Phi_{RR})_{ii} &=-\sum_{j\in\th_{R}\setminus\{i\}}  \frac{t_j}{t_i}  (\Phi_{RR})_{ij}  +  O(q)  \,.
\end{align*}
where $\th^{*}_{L}=\th_{L}\cup\{*\}$. This is precisely the form required for the expected channel.

With our choices corresponding to a positive helicity intermediate state on $\Sigma_L$, it is natural to remove a row and column corresponding to a graviton on $\Sigma_R$ when computing the original $\mathrm{det}'\Phi$. As $q\rightarrow 0$, one finds
\be\label{Phifac}
\mathrm{det}^{\prime} \Phi = \, \sqrt{q} \, t_\ast^2 \, \mathrm{det}^{\prime} \Phi_{L}\,\mathrm{det}^{\prime }\Phi_{R}+O(q)\,,
\ee
where $\Phi_L$ is the $(|\th^*_{L}|+1)\times(|\th^*_{L}|+1)$ matrix appropriate for $\Sigma_L$ with the positive helicity intermediate state $*$ at $z_a\in\Sigma_L$, and $\Phi_R$ is the $|\th_{R}|\times|\th_{R}|$ matrix appropriate for $\Sigma_R$.

The story for $\tilde{\Phi}$ proceeds in a similar fashion. By introducing 
\be\label{int2}
1=\int\d^{2}\lambda_{*}\,\delta^{2}\!\left(\lambda_{*}-t_{*}\lambda(z_a)\right)\,,
\ee
in \eqref{EYMform}, with
\be\label{int2*}
\lambda(z_a):=\sum_{k\in\tth_{L}\cup\ttg}\tilde{t}_{k}\,\lambda_{k}\,S(a,k)\,,
\ee
the diagonal entries of $\tilde{\Phi}_{LL}$ and $\tilde{\Phi}_{RR}$ become
\begin{align*}
(\tilde{\Phi}_{LL})_{kk}& =-\sum_{l\in\tth_{L}\setminus\{k\}} \frac{\tilde t _l}{ \, \tilde t _k} \, \tilde{\Phi}_{kl}- \frac{\tilde t _r}{\, \tilde t _ k} \,  \tilde{\Phi}_{k\ttg} + O(q) \,, \\
(\tilde{\Phi}_{RR})_{kk}& =-\sum_{l\in\tth^{*}_{R}\setminus\{k\}}  \frac{\tilde t _l}{ \, \tilde t _k} \, \tilde{\Phi}_{kl} +O(q) \,,
\end{align*}
where $\tth^{*}_{R}=\tth_{R}\cup\{*\}$. With this corresponding to a negative helicity intermediate state on $\Sigma_R$, we take $\mathrm{det}^{\prime}\tilde{\Phi}$ by eliminating a row and column for a graviton on $\Sigma_L$, to obtain the desired factorization:
\be\label{tPhifac}
\mathrm{det}^{\prime}\tilde{\Phi}= \sqrt{q} \, \tilde t_\ast ^2 ~ \mathrm{det}^{\prime} \tilde{\Phi}_{L}\,\mathrm{det}^{\prime}\tilde{\Phi}_{R}+O(q)\,.
\ee
This shows that the reduced determinants factorize correctly and yield a factor of $q \, t_\ast^2 \tilde t _\ast^2$, while also introducing the wave functions for an intermediate on-shell state.

Next we examine the behaviour of the external wave functions. On the support of \eqref{int1}, \eqref{int2}, the arguments of the delta functions for the external states involve
\be\label{dlam}
t_i \, \lambda(z_i)=\left\{\begin{array}{ll}
                           \sum_{k\in\tth_{L}\cup\ttg}t_i \tilde{t}_{k}\,\lambda_{k}\,S(i,k) +O(q) \:\: & \mbox{for } z_i \in\Sigma_{L} \\
                           \sum_{k\in\tth^{*}_{R}} t_i \tilde{t}_{k}\,\lambda_{k}\,S(i,k) \:\: & \mbox{for } z_i \in\Sigma_{R}
                          \end{array}\right.\,,
\ee
\be\label{dtlam}
\tilde t_ k \, \tilde{\lambda}(z_k)=\left\{\begin{array}{ll}
                           \sum_{i\in\th^{*}_{L}\cup\tg} t_{i} \tilde t _k \,\tilde{\lambda}_{i}\,S(k,i)  \:\: & \mbox{for } z_k \in\Sigma_{L} \\
                           \sum_{i\in\th_{R}\cup} t_{i} \tilde t _k \,\tilde{\lambda}_{i}\,S(k,i) +O(q)  \:\: & \mbox{for } z_k \in\Sigma_{R}
                          \end{array}\right.\,.
\ee
in the degenerate limit $q\rightarrow 0$.

\medskip

As $q\rightarrow0$ a scaling argument based on the local model \eqref{locmod} dictates how the measure on the moduli space of $\Sigma$ factorizes into measures on the moduli spaces of $\Sigma_L$ and $\Sigma_R$ (\textit{cf.} \cite{Polchinski:1988jq,Witten:2012bh}). Combining all of our observations up to this point, the formula \eqref{EYMform} looks like
\begin{multline}
\int \frac{\d q}{q^2}\,\frac{\d^{2}\lambda_{*}\,\d^{2}\tilde{\lambda}_{*}}{(\mathrm{vol}\;\C^{*})^2} \frac{ \d z_{a}\,\d z_{b}}{(\mathrm{vol}\;\SL(2,\C))^2}\, \delta\!\left(1 - \frac{1}{t_* \tilde{t}_*}\frac{1}{\sqrt{\d z_{a}\,\d z_{b}}}\right) ~~ \\ \times q \, t_\ast ^2 \,\tilde{t}_\ast^2 ~~  \mathrm{det}^{\prime} \Phi_{L}\,  \mathrm{det}^{\prime} \tilde{\Phi}_{L}\, ~ \mathrm{det}^{\prime} \Phi_{R}\,  \mathrm{det}^{\prime}\tilde{\Phi}_{R}\; \rPT \\
\frac{\d t_{*}}{t_* ^3}\,\delta^{2}\!\left(\lambda_{*}-t_{*}\lambda(z_a)\right)\prod_{i\in\th_{L}\cup\tg_{L}}\frac{\d t_i}{t_{i}^{|2h|-1}}\,\delta^{2}\!\left(\lambda_{i}-t_{i}\lambda(z_i)\right)\, \prod_{k\in\tth_{L}\cup\ttg_{L}}\frac{\d\tilde{t}_{k}}{\tilde{t}_{k}^{|2h|-1}}\,\delta^{2}\!\left(\tilde{\lambda}_{k}-\tilde{t}_{k}\tilde{\lambda}(z_k)\right) \\
\frac{\d\tilde{t}_*}{\tilde{t}_* ^3}\, \delta^{2}\!\left(\tilde{\lambda}_{*}-\tilde{t}_{*}\tilde{\lambda}(z_b)\right)\prod_{j\in\th_{R} }\frac{\d t_j}{t_{j}^{|2h|-1}}\,\delta^{2}\!\left(\lambda_{j}-t_{j}\lambda(z_j)\right)\, \prod_{l\in\tth_{R}}\frac{\d\tilde{t}_{l}}{\tilde{t}_{l}^{|2h|-1}}\,\delta^{2}\!\left(\tilde{\lambda}_{l}-\tilde{t}_{l}\tilde{\lambda}(z_l)\right)\,,
\end{multline}
up to $O(q)$, where we have reverted to the original, unrescaled matrices $\Phi , \tilde \Phi$. 

The delta functions for $\lambda_{*}$ and $\tilde{\lambda}_{*}$ are naturally incorporated into products over $\prod_{i\in\th^{*}_{L}\cup\tg}$ and $\prod_{l\in\tth^{*}_{R}}$ as gravitons. Finally, trading the delta function of \eqref{const} for an additional $(\mathrm{vol}\,\C^*)^{-1}$, we are left with
\be\label{fact1*}
\int \frac{\d q}{q} \,\frac{\d^{2}\lambda_{*}\,\d^{2}\tilde{\lambda}_{*}}{\mathrm{vol}\;\C^{*}}\,\left(\cM^{+}_{L}\,\cM^{-}_{R}+O(q)\right)\,,
\ee
where
\begin{equation*}
 \cM^{+}_{L}\!=\!\int\!\frac{\mathrm{det}^{\prime}\Phi_{L}\mathrm{det}^{\prime}\tilde{\Phi}_{L}}{\mathrm{vol}\;\GL(2,\C)} ~~\rPT \!\!\prod_{i\in\th^{*}_{L}\cup\tg}\!\frac{\d t_i}{t_{i}^{|2h|-1}}\delta^{2}\!\left(\lambda_{i}-t_{i}\lambda(z_i)\right)\!\! \prod_{k\in\tth_{L}\cup\ttg}\!\frac{\d\tilde{t}_{k}}{\tilde{t}_{k}^{|2h|-1}}\delta^{2}\!\left(\tilde{\lambda}_{k}-\tilde{t}_{k}\tilde{\lambda}(z_k)\right)
\end{equation*}
and similarly for $\cM_{R}^{-}$.

All that remains to show is that extracting the residue at $q=0$ corresponds to setting the momentum flowing from $\Sigma_L$ to $\Sigma_R$ on-shell. To this end, notice that the various delta functions in \eqref{fact1*} imply that
\begin{equation*}
 \sum_{i\in\th_{L}\cup\tg}\lambda_{i}\tilde{\lambda}_{i}\,\,+\sum_{k\in\tth_{L}\cup\ttg} \lambda_{k}\tilde{\lambda}_{k}= \lambda_{*}\tilde{\lambda}_{*}+O(q)\,.
\end{equation*}
In addition to enforcing momentum conservation on each side of the cut, this reveals the intermediate particle as manifestly on-shell. Thus, taking the residue at $q=0$ results in the $h=+$ graviton exchange term in the factorization expression \eqref{bfp}. The $h=-$ term is obtained in a similar fashion, by choosing the other sign in  \eqref{t-scaling}.

\medskip

The second possible degeneration, which disturbs the colour structure of the amplitude via a gluon exchange, follows in much the same way, so we will be more brief in its description. In this case the external gluons are split between $\Sigma_{L}$ and $\Sigma_{R}$ in the $q\rightarrow0$ limit. The behaviour of the matrices $\Phi$ and $\tilde{\Phi}$ under the degeneration is the same as before; it is now convenient to eliminate the row and column corresponding to the trace in both $\mathrm{det}^{\prime}\Phi$ and $\mathrm{det}^{\prime}\tilde{\Phi}$. 

Once again, after inserting the delta functions \eqref{int1}, \eqref{int2}, the reduced determinants factorize as \eqref{Phifac}, \eqref{tPhifac} with the intermediate particle being incorporated into $\tg^{*}_{L}=\tg_{L}\cup\{*\}$ on $\Sigma_L$ and $\ttg^{*}_{R}=\ttg_{R}\cup\{*\}$ on $\Sigma_R$. However, the factor of $q\,t_{*}^2 \tilde{t}_{*}^2$ is entirely absorbed by the scaling of the gluon representatives $t_r^2$, $\tilde{t}_{s}^2$ appearing in \eqref{eqn:rescaled-matrices}. 

From \eqref{Szd} it is easy to see that the Parke-Taylor factor behaves as
\be\label{PTfact}
\rPT = q \,\, t_\ast^2 \tilde t_\ast ^2 ~ \,\rPT_{ {L}^{a}}\,\rPT_{ {R}^{b}}+O(q^{3/2})\,,
\ee
where $\rPT_{\gamma_{L}^{a}}$ denotes the Parke-Taylor factor for those gluons located on $\Sigma_L$ with the point $z_a$ inserted in the cyclic ordering precisely where the original colour trace is broken by the degeneration. The measures for $t_{*}, \tilde{t}_{*}$ appearing in \eqref{const} are automatically appropriate for gluons, so we obtain the correct products over delta functions for factorization. Taking the residue of the resulting $\d q/q$ measure corresponds to the $h=+$ gluon exchange term in the factorization expression \eqref{bfp}.

\medskip

Factorization for the supersymmetric amplitude \eqref{sEYM} is established by looking at how the exponential encoding supersymmetry behaves in the $q\rightarrow0$ limit. For concreteness we pick the case of \eqref{t-scaling} where $t_i \sim q^{1/4} $ on $\Sigma _L$ and the other case follows analogously, as usual. To begin notice that the exponent splits as
\begin{equation}
 \sum_{\substack{i\in\th_{L}\cup\tg_{L} \\ k\in\tth_{L}\cup\ttg_{L}}}   \!\!\!\!\!    t_{i}\tilde{t}_{k}\,\tilde{\eta}_{i} \! \cdot \! \eta_{k}\,S(k,i) +\sum_{\substack{i\in\th_{R}\cup\tg_{R} \\ k\in\tth_{R}\cup\ttg_{R}}}  \!\!\!\!\!    t_{i}\tilde{t}_{k}\,\tilde{\eta}_{i} \! \cdot \!  \eta_{k}\,S(k,i)   +\sum_{\substack{i\in\th_{R}\cup\tg_{R} \\ k\in\tth_{L}\cup\ttg_{L}}}  \!\!\!\!\!   t_{i}\tilde{t}_{k}\, t_\ast \tilde t _\ast \, \tilde{\eta}_{i} \! \cdot \! \eta_{k}\,S(k,a) S(b,i)  +O(q)\,.
\end{equation}
The first two terms are clearly appropriate for $\Sigma _L$ and $\Sigma _R$, respectively, while the third term accounts for the new intermediate particle. In fact, a simple calculation reveals that the exponential of this third term can be written
\begin{align*}
& ~ \exp  \left( \sum_{\substack{i\in\th_{R}\cup\tg_{R} \\ k\in\tth_{L}\cup\ttg_{L}}}     t_{i}\tilde{t}_{k}\, t_\ast \tilde t _\ast \, \tilde{\eta}_{i} \! \cdot \! \eta_{k}\,S(k,a) S(b,i)   \right) \\
&= \int \d^ \cN \eta_\ast  \d^ \cN \tilde \eta_\ast    ~~ \e^{\eta _\ast \cdot \tilde \eta _ \ast} \, ~ \exp \left(   \sum_{  k\in\tth_{L}\cup\ttg_{L}}  \!\!\!   t_{\ast}\tilde{t}_{k}\,   \tilde{\eta}_{\ast} \! \cdot \! \eta_{k}\,S(k,a)   +    \sum_{i\in\th_{R}\cup\tg_{R}  }  \!\!\!   t_{i}   \tilde t _\ast \, \tilde{\eta}_{i} \! \cdot \! \eta_\ast \, S(b,i)    \right) ~,
\end{align*}
on the support of the same delta functions used in the purely bosonic calculation. 

In summary, the exponential encoding supersymmetry factorizes, with the arguments on $\Sigma_L$, $\Sigma_R$ becoming
\begin{equation*}
 \sum_{\substack{i\in\th^{*}_{L}\cup\tg_{L} \\ k\in\tth_{L}\cup\ttg_{L}}}t_{i}\tilde{t}_{k}\,  ~ \tilde{\eta}_{i}\cdot\eta_{k}\,S(k,i)  \qquad \text{and} \qquad \sum_{\substack{i\in\th_{R}\cup\tg_{R} \\ k\in\tth_{R} ^\ast \cup\ttg_{R}}}t_{i}\tilde{t}_{k}\, ~ \tilde{\eta}_{i}\cdot\eta_{k}\,S(k,i) \,, 
\end{equation*}
respectively (without altering the simple pole in $q$). Combined with our previous arguments, this leads to the appropriate on-shell superspace measure
\be\label{sfact}
\int \frac{\d q}{q}\,\frac{\d^{2|\cN}\lambda_{*}\,\d^{2|\cN}\tilde{\lambda}_{*}}{\mathrm{vol}\;\C^{*}}\,\,\e^{\eta_{*}\cdot\tilde{\eta}_{*}}\left(\cM^{+}_{L}\,\cM^{-}_{R}+O(q)\right)\,.
\ee
So factorization of the super-amplitude follows straightforwardly from factorization of the bosonic formula.


\section{Conclusions}

We have presented a new formula for all single trace tree amplitudes in supersymmetric Einstein Yang-Mills theory in four dimensions, written in terms of on-shell superspace. The formula was shown to reproduce the correct three-point amplitudes and to factorize appropriately in both gravitational and coloured channels. The considerations at the beginning of Section~\ref{form} also place strong constraints on the form of general multitrace EYM amplitudes. It would be interesting to investigate these further.

In the purely gravitational sector, the formula reduces to the representation of amplitudes given in~\cite{Cachazo:2012kg,Geyer:2014fka}. These formulae are known to be the output of a twistor-string theory for maximal supergravity in $d=4$~\cite{Skinner:2013xp} and it is natural to wonder whether there is a modification of this theory that describes sEYM. In this regard, we note that the coupling of the gluons to the gravitons in $\Phi,\tilde{\Phi}$, together with the Parke-Taylor factors, may be generated by inserting operators
\begin{equation*}
\tr\left(\bar{D}^{-1}\,\delta^{2}(\gamma)\,\cA\,\bar{D}^{-1}\,\delta^{2}(\tilde{\gamma})\,\tilde{\cA}\right)\,, \qquad \mbox{and} \qquad \tr\left(\bar{D}^{-1}\,\cO_{\cA}\,\bar{D}^{-1}\,\tilde{\cO}_{\tilde{\cA}}\right)\,.
\end{equation*}
Here $\bar{D}=\dbar+\cA(Z)+\tilde{\cA}(W)$ and
\begin{equation*}
\cO_{\cA}=\left[W,\frac{\partial\cA}{\partial Z}\right]+ \left[\bar{\rho}, \rho^{K}\frac{\partial^{2} \cA}{\partial Z \partial Z^{K}}\right]\,,\qquad \tilde{\cO}_{\tilde{\cA}}=\left\la Z,\frac{\partial\tilde{\cA}}{\partial W}\right\ra +\left\la \rho, \bar{\rho}_{K}\frac{\partial^{2}\tilde{\cA}}{\partial W \partial W_{K}}\right\ra\,,
\end{equation*}
where $\cA$ and $\tilde{\cA}$ are gluon wavefunctions. It seems likely that the RSVW formula for tree amplitudes in sYM are best interpreted as the single trace sector of a twistor string for sEYM. Such a theory would also enable the use of worldsheet factorization arguments to streamline the calculations above~\cite{Adamo:2013tca}.

\acknowledgments

The work of TA is supported by a Title A Research Fellowship at St. John's College, Cambridge. The work of EC is supported in part by the Cambridge Commonwealth, European and International Trust. KR and DS are supported in part by a Marie Curie Career Integration Grant (FP/2007-2013/631289).

\bibliography{EYM_JHEP}
\bibliographystyle{JHEP}

\end{document}